# Observation of single-quantum vortex splitting in the $Ba_{1-x}K_xFe_2As_2$ superconductor

Q. Z. Zhou[1,*], B. R. Chen[1,*], B. K. Xiang[1,*], I. Timoshuk[2,3], J. Garaud[4], Y. Li[5], K. Y. Liang[1], Q. S. He[1], Z. J. Li[1], P. H. Zhang[1], K. Z. Yao[1], H. X. Yao[1], E. Babaev[2,3,†], V. Grinenko[5,†], Y. H. Wang[1,6,†]

1. State Key Laboratory of Surface Physics and Department of Physics, Fudan University, Shanghai 200433, China
2. Department of Physics, Royal Institute of Technology, SE-10691 Stockholm, Sweden
3. Wallenberg Initiative Materials Science for Sustainability Department of Physics, Royal Institute of Technology, SE-10691 Stockholm, Sweden
4. Institut Denis Poisson CNRS/UMR 7013, Universite de Tours, 37200 France
5. Tsung-Dao Lee Institute and School of Physics and Astronomy, Shanghai Jiao Tong University, Shanghai 201210, China
6. Shanghai Research Center for Quantum Sciences, Shanghai 201315, China

\* These authors contributed equally to this work.
† Email address: babaev@kth.se; vadim.grinenko@sjtu.edu.cn; wangyhv@fudan.edu.cn

**Abstract**

**Since their theoretical discovery over a half-century ago, vortices observed in bulk superconductors have carried a quantized value of magnetic flux determined only by fundamental constants. A recent experiment reported 'unquantized' quantum vortices carrying the same fraction of flux quantum in $Ba_{0.23}K_{0.77}Fe_2As_2$ in a small temperature range below its superconducting critical temperature ($T_C$). Here, we use scanning superconducting quantum interference device (sSQUID) microscopy with improved sensitivity to investigate the genesis of fractional vortices in $Ba_{0.23}K_{0.77}Fe_2As_2$. We report the direct observation of a single-flux quantum vortex splitting into two different fractions with increasing temperature. The flux of the two fractions has opposite dependence on temperature, while the total flux sums up to one flux quantum despite their spatial separation. Overall, our study shows the existence of different fractional vortices and their stability in temperature ranging from 0.1 to 0.99 $T_C$. Besides**

**the implications of this observation for the fundamental question of quantum vorticity, the discovery of these objects paves the way for the new platform for anyon quasiparticles and applications for fractional fluxonics.**

## Introduction

Magnetic flux penetrates a superconductor by forming vortices that carry quantized magnetic flux $\Phi = N\Phi_0$, where $\Phi_0 = \frac{h}{2e}$ is the flux quantum, and $N$ is an integer[1–4]. For decades, that was a universally observed quantization condition for various superconductors irrespective of their composition, critical temperature, or degree of disorder[5]. As a superconducting state breaks $U(1)$ symmetry, the robustness of London's flux quantization is commonly considered to be deeply rooted in the $U(1)$ symmetry and topology of the system. Flux quantization thus serves as one of the pioneering examples of a topological charge concept in physics. More general defects were theoretically considered in unconventional spin-triplet superconductors, using the generalization of these symmetry and topology principles. It was theoretically expected that chiral $p$-wave superconductors allow the formation of vortices with flux $\Phi = \Phi_0/2$ [6]. However, the half-quantum flux was so far only observed in mesoscopic rings as configurations stabilized by geometric effects[7,8], rather than in the form of true quantum vortices in bulk systems.

Usually, the term quantum implies the strict lower bound on the possible value of a quantity. A different argument was advanced that vorticity in superconductors could arise outside of the usual symmetry and topology arguments[9]. It was suggested that vortices of different origins could carry an arbitrarily small fraction of magnetic flux quantum. These vortices were coined recently as 'unquantized' since the value of the flux fraction depends on material parameters. Such vortices were predicted to occur in multiband materials[9]. They can originate from a phase winding in a complex field $\psi_i$ that describes superconductivity originating in an $i$-th electronic band. Usually, fractional vortices are denoted as $(n_1, n_2, n_3)$, where $n_j$ is a phase winding number in the corresponding band. The one quanta vortex in this notation is denoted as (1, 1, 1). Then a vortex (1, 0, 0) carries the following fraction of the flux quantum: $\Phi_1 = \Phi_0|\psi_1|^2/\sum|\psi_i|^2$. Recently, numerical solutions for fractional vortices were obtained in fully self-consistent microscopic models[10,11]. Furthermore, an important implication of unquantized flux is that the cores of such vortices realize the bound state of electronic degrees of freedom to fractional flux, which is a long thought-after composite 'particle-like' object that fulfils the conditions[12] of obeying anyonic statistics.

Recently, unquantized vortices were observed in Ba$_{1-x}$K$_x$Fe$_2$As$_2$ [10], which has at least three superconducting bands[13]. At $x \approx 0.77$, it broke time-reversal symmetry[14–16], which was consistent with $s + is$ state[17–20]. Only a single type $\Phi_1$ of a fractional vortex was observed in Ba$_{0.23}$K$_{0.77}$Fe$_2$As$_2$ in the temperature ranges from 9 K up to $T_c \approx 11$ K using sSQUID microscopy [10]. The fraction $\Phi_1$ was monotonically decreasing with temperature so that below $T < 9$ K, it was no longer possible to resolve fractional vortices. The missing counterpart fractions $\Phi_{j \neq 1}$ was a puzzling aspect of the experiment. Interestingly, work in parallel to the current one reports fractionalization of the Abrikosov vortex core singularities at the nanoscale in a different but related material using scanning tunneling spectroscopy[21]. However, the values of the quantum flux fractions cannot be deduced from the fractionalization of the core, as it requires magnetometry measurements.

Here, we report the first direct observation of the fractionalization process of magnetic flux quantum in Ba$_{1-x}$K$_x$Fe$_2$As$_2$. That allows us to observe the previously missing counterpart fraction of the flux quantum. We find evidence that the fractions have opposite temperature dependence. Namely, in addition to the fraction that increases with temperature, as reported in[10], we observe the fraction that decreases with increasing temperature. Despite their spatial separations, they vary in a correlated 'seesaw'-like way, adding up to a single flux quantum. Our experimental data along with the theoretical analysis establish that the unquantized vortices are associated with the different phase windings sustained in different bands.

**Integer vortex splitting into fractions with increasing temperature**

In order to detect fractional vortices away from $T_C$, we need enhanced sensitivity with sSQUID microscopy. In this work, the higher sensitivity is achieved by mounting the nano-SQUID on a tuning fork (Fig. 1a). We are able to obtain a demodulated AC flux signal $\Phi_{AC}$ due to the vibration of the tuning fork[22–25]. Like in a vibration sample magnetometer, $\Phi_{AC}$ is proportional to the DC magnetic signal $\Phi_{DC}$ when the vibration amplitude is small. Since we use a flux-locked loop to obtain $\Phi_{DC}$, it provides an absolute value for the flux[23,26]. $\Phi_{AC}$ is more sensitive to small variations of the flux signal. Both are essential for detecting fractional vortices away from $T_c$ and assigning absolute flux values for them.

We use this capability to measure a $Ba_{0.23}K_{0.77}Fe_2As_2$ sample (Fig. 1b), with nearly the same composition as the one where one kind of fractional vortices was observed near $T_C$[10]. After cooling under zero field to 3.18 K, the sample shows remanent integer vortices and antivortices in both $\Phi_{DC}$ (Fig. 1c) and $\Phi_{AC}$ (Fig. 1d) images. $\Phi_{DC}$ and $\Phi_{AC}$ images of a different area (Figs. 1e and f) show similar patterns overall, while its susceptometry image is also similar (Figs. 1c and e, insets). Unsurprisingly, the integer vortices at these two different areas (Figs. 1c-f, dashed boxes) located away from other vortices are not too different either in terms of their shape and total flux.

Fractional vortices are always more energetically expensive[9], and their formation can be a relatively rare effect. We find that the two seemingly similar integer vortices behave quite differently when temperature increases. The integer vortex in Figures 1c and d does not change much in shape and total flux with increasing temperature other than getting larger due to increasing penetration depth (Extended Data Fig. 1). Therefore, we assign it as a conventional integer vortex (CV). However, the integer vortex in the other area (Figs. 1e and f) behaves very differently with increasing temperature (Fig. 2). Although there is no qualitative difference in $\Phi_{DC}$ and $\Phi_{AC}$ images from the base temperature to 3.94 K (Figs. 2(i-ii)), the apparent changes start to appear at elevated temperatures. The upper part of the vortex becomes slightly pointed at 5.6 K and 8.1 K, which is more obvious in $\Phi_{AC}$ (Figs. 2b(iii-iv)). The extruded upper part of the vortex clearly splits off at 9.83 K into a weak fractional vortex (FV) and leaves the lower part as a stronger FV with the same sign (Fig. 2b(v)). The splitting continues at 10.52 K (Fig. 2b(vi)), and both the intensity and shape of the FVs evolve with temperature.

The upper FV gains more flux, while the lower one loses intensity with increasing temperature. Both FV becomes more distinctive in both $\Phi_{DC}$ (Figs. 2a(vi-x)) and $\Phi_{AC}$ (Figs. 2b(vi-x)), suggesting their separation is getting larger. The shapes of the upper and lower fractions are also distinctively different at this stage. The lower one is more circular, while the upper fraction is smaller and has a pointed top, resulting in a gourd-like profile. The gourd becomes more top-heavy at even higher temperatures (Figs. 2b(xi-xv)). Above 10.9 K, the flux in the scanned area is affected by nearby vortices and antivortices, yet the fractionalization remains clearly visible. At 10.93 K, the upper FV becomes stronger than the lower one (Fig. 2xi). The upper FV keeps getting stronger and the lower FV weaker till 10.99 K (Fig. 2xii). At temperatures within 0.15 K of $T_C$, both become broader and show less contrast (Figs. 2a(xiii-xv)), consistent with the diverging penetration depth at this temperature regime. We have also found an instance of an integer vortex of $-\Phi_0$ splitting into two negative FVs

(Extended Data Fig. 3). In that case, it is the lower fraction that splits off (Extended Data Fig. 3b(iv)). The lower fraction appears weak at first and then gets stronger with increasing temperature, while the upper fraction behaves the opposite way. When fractions remain within the scanned area, and the area is free from antivortices, the total flux adds to a single flux quantum irrespective of temperature in agreement with the multiband model[9].

In order to gain a quantitative understanding of the vortex-splitting process, we analyze the flux redistribution between the split FVs. We take linecuts of the split vortices' $\Phi_{AC}$ along their central y-axis (Fig. 3a). The decreasing height of the lower FV peak and increasing height of the upper one with increasing temperature indicates the transfer of magnetic flux from the lower to the upper FV. The separation between the peaks increases with temperature and reaches a maximum of 10.87 K (Fig. 3a, grass green). The diamagnetic susceptibility as a function of temperature (Fig. 4b, black diamonds and right axis) and $T_C$ = 11.2 K is similar to the sample with similar doping studied in[10,15].

Keeping the sequence of vortex splitting in mind, we obtain the temperature dependence of the magnetic flux of the upper and lower FVs by integrating the magnetic field over their respective region (Fig. 3b). The fractional vortices should be well isolated and fit within the scanning area in order to compare with the theoretical fractional quantization condition[9] and experiment on isolated vortices[10]. Such condition is approximately met up to ~ 10.87 K only. Flux values at higher temperatures cannot be compared with those of isolated vortices due to the influence of the closely located antivortices.

The flux of the upper fraction increases with temperature (Fig. 3b, red triangles) while the lower fraction decreases (Fig. 3b, blue triangles) from the base temperature up to 10.93 K. The change of flux with increasing temperature is slower in the low-temperature range but more dramatic at temperatures above 0.8 $T_C$. Significantly, their sum equals one unit of flux quantum $\Phi_0$ (within an error of ~10%) from base temperature up to 0.965 $T_C$ (Fig. 3b, black circles). For the splitting of the vortex with the opposite vorticity (Extended Data Fig. 3), the total flux of the upper and lower FV also equal $\Phi_0$ up to the same temperature (Extended Data Fig. 4b). This observation strongly suggests that the flux is transferring from the lower FV into the upper FV and the two fractions remain correlated after they split. The flux values of the fraction that grows with temperature are qualitatively consistent with those observed in Iguchi et al[10]. In this work, we identified another type of fractional vortex whose fraction reduces with increasing temperature (Fig. 3b, blue triangles). That type of vortex can have phase winding in one or several other bands.

We have also found instances of fractional vortices with no fractional vortex partner nearby. We observe that even at the base temperature with fractions different from the ones in Figure 2 (Fig. 4 and Extended Data Fig. 2b). The total flux of such isolated fractional vortices also increases with temperature (Fig. 3b, green circles and orange squares). However, at comparable temperatures, they seem to have twice as large flux overall than the upper FV and approximately four times larger than the FV reported in Iguchi *et al.*[10] (Fig. 3b, pink diamonds). However, their flux is smaller than the lower FV (Fig. 3b, blue triangles). That suggests that some of the fractional vortices that we observe have windings in several bands or represent clusters of pinned fractional vortices. The difference in fluxes between the observed FVs in this work and those in Iguchi *et al.*[10] may be related to a slight difference in the samples' doping. Several fractional vortices with separated cores on the nanoscale may exist, according to the results of the parallel scanning tunneling microscopy studies[18].

Having shown the splitting of a CV into two major fractions with opposite temperature dependence over a large temperature range, we now examine more complex fractionalization close to $T_C$ (Figs. 2xi-xv). The material is known to have strong fluctuations[15], and hence, fluctuation-generated mobile vortices are to be expected. At 10.93 K, a fractional antivortex appears on the upper left side of the upper FV (Fig. 2xi). The fractional antivortex becomes stronger and larger with further increasing temperature (Figs. 2xii-xv). In the upper right corner is a sign of another weaker FV at 11.05 K (Fig. 2xiii).

The total flux of the upper FV (Fig. 3c, red triangles) decreases with temperature within 1.5% of $T_C$. This contrasts with the monotonic increase of isolated FVs until $T_C$ [10] (Fig. 3b, solid diamonds). The temperature turning point of the flux is coincident with the sudden downward shift of the upper FV (Fig. 3a). The appearance of the additional fractional vortex and antivortex coincident with the decreasing flux of the upper FV demonstrates more complex fluctuation-affected processes close to $T_C$. The allowed fluctuation processes involve mergers with fluctuation-generated composite antivortices (-1, -1, -1), which convert some fractional vortices (e.g. (0, 1, 0)) to fractional antivortices (e.g. (-1, 0, -1))[27]. Furthermore, if the larger fraction represents a composite fractional vortex (0, 1, 1) at elevated temperature, it can fractionalize further to (0, 1, 0) + (0, 0, 1). We note that some fractions may have left the scanning area in our images obtained close to $T_C$. As this represents an interesting regime connected with vortex-driven phase transtions[15], it calls for further magnetometry studies.

**Analysis of temperature dependence**

According to the theoretical models[9,28], fractionalization is associated with phase windings in different bands. However, the very strong and nontrivial temperature dependence of flux quantum fractions is not present in the simplest models[9,28]. In this section, we analyze what is required to obtain such a nontrivial temperature dependence and whether the microscopic conditions are consistent with the band structure of $Ba_{1-x}K_xFe_2As_2$. Several nontrivial constraints should be met to be consistent with other experimental facts known about $Ba_{1-x}K_xFe_2As_2$: (i) at $x \approx 0.77$, the critical temperature $T_C^{Z2}$ at which the time-reversal symmetry is broken reaches maximum[15]. This requires three gaps of similar magnitude according to standard $s + is$ models [17–20]. At the same time, a model with three fully symmetric bands has vortices that carry $\approx 1/3$ of $\Phi_0$ independent of temperature[9].

We consider a minimal three-band microscopic model of a clean superconductor to understand the temperature dependence. The full derivation is detailed in the supplementary materials. Within the quasiclassical approximation, the solutions of the Eilenberger equations for the quasiclassical propagators are expressed as an expansion by powers of the gap functions amplitudes $\Delta_a$ and of their gradients. Subsequently, the Ginzburg-Landau equations are obtained by inserting such an expansion in the self-consistency equation

$$\Delta_a(p,r) = 2\pi T \sum_{n,p',b} \lambda_{ab}(p,p') f_b(p,r,\omega_n)$$

Here, the parameters $p$ and $p'$ are the momenta that run over the Fermi surfaces, $\lambda_{ab}$ are the components of the coupling potential matrix $\widehat{\Lambda}$, and $\omega_n$ are the fermionic Matsubara frequencies.

Overall, all the parameters of the resulting Ginzburg-Landau equations, self-consistently determined within the quasiclassical approximation, depend on the pairing potential matrix and the parameters of each Fermi surface (the partial densities of states $\nu_a$ and the Fermi velocities $v_F^a$). The first constraint of the parameters in our case is that they should be chosen to qualitatively reflect the doping level that gives the maximal (or near maximal) critical temperature $T_C^{Z2}$ for the breakdown of time-reversal symmetry. This occurs when three gaps have approximately equal amplitudes, maximizing the frustration of interband Josephson coupling terms (this frustration is responsible for the breakdown of time-reversal symmetry in $s + is$ superconductors[17–19]). We consider the case of an intraband-dominated pairing with repulsion. Note that the Ginzburg-Landau-based analysis is not quantitatively valid at low temperatures, where the expansion fails. Also, since it is based on mean-field approximation, it is not valid close to critical temperature $T_C < T_C^{MF}$ due to strong fluctuation effects in $Ba_{0.23}K_{0.77}Fe_2As_2$ [15]. Hence, we only consider the range of temperatures where these approximations can be justifiable (supplementary materials). This model produces a temperature dependence of the different flux fractions in qualitative agreement with the experiment under several conditions (Figs. 3d and e).

**Isolated fractional vortex far from critical temperature**

The sensitivity of the experimental setup allows for the study of the comparative morphology of the fractional and integer vortices. It was shown earlier that fractional vortices in multiband systems have, in general, different localization of magnetic field[29]. As the two fractions of a split vortex reported above are still relatively close, we turn to the isolated FV to investigate its intrinsic geometric shape. To compare side-by-side, we show the isolated CV (Fig. 4a, also in Fig. 1d) and the FV (Fig. 4b). Because the point-spread-function of the nano-SQUID slightly distorts the $\Phi_{AC}$ image, the CV does not appear utterly circular in it. However, the linecuts along the horizontal and vertical direction through its centre show similar widths (Fig. 4c, red and orange). The FV, however, clearly breaks axial symmetry. It shows a much more elongated shape with different widths in the linecuts along the two orthogonal directions (Fig. 4c, blue and purple), which can also be seen from the normalized linecuts (Fig. 4d). The $\Phi_{AC}$ images of the isolated FV at various temperatures (Fig. 4e) show that its broken axial symmetry is getting stronger with increasing temperature. Aspect ratios of the FV obtained from the images exhibit a rising trend with temperature (Fig. 4f, blue circles), in contrast to the unit constant exhibited by the CV (Fig. 4f, orange squares).

There could be several reasons for this elongation. One reason could be that two (1, 0, 0) vortices are pinned nearby. However, such vortices strongly repel each other[9,30,31]. It may represent a pair of (1, 0, 0) and (0, 1, 0) vortices which interact attractively but may split in the presence of a domain wall[32]. However, it can also be an elementary fractional vortex that spontaneously breaks axial symmetry. Since that is observed far below critical temperature, checking the breakdown of axial symmetry of fractional vortices requires in general, a fully microscopic calculation. To that end, we performed calculations in the three-band Bogoliubov-de-Gennes model; details are given in Supplementary Materials. The fractional vortex breaking axial symmetry in an $s + is$ superconductor, according to our numerical calculation (Figs. 4g and h), is present even in this minimal model. The breaking of axial symmetry can be temperature-dependent.

Our flux sensitivity also allows us to establish that fractional vortices exist far below the critical temperature, and their magnetic flux fraction varies much slower as a function of temperature far below $T_C$. We find an unusual regime near critical temperature where a dense ensemble of vortices and antivortices appear in the scanned area. This is an indication of beyond-mean-field effects[17]. Besides the direct observation of flux quantum fractionalization, our results enable new fluxomic schemes where information density can be increased by using base-$N$ schemes enabled by encoding information not only in inter vortices but also in fractions of magnetic flux quantum. Furthermore, our observation is important for the possibility of realizing anyons in this type of platform.

**Acknowledgement**
We would like to acknowledge support by National Natural Science Foundation of China (Grant No. 12150003), Shanghai Municipal Science and Technology Major Project (Grant No. 2019SHZDZX01) and National Key R&D Program of China (Grant No. 2021YFA1400100). EB and IT were supported by the Swedish Research Council Grants 2022-04763, by Olle Engkvists Stiftelse, and the Wallenberg Initiative Materials Science for Sustainability (WISE) funded by the Knut and Alice Wallenberg Foundation. VG acknowledges support from the National Natural Science Foundation of China (No. 1237040280).

**Data availability**
The data that support the findings of this work are available from the corresponding authors upon reasonable request.

# Figures

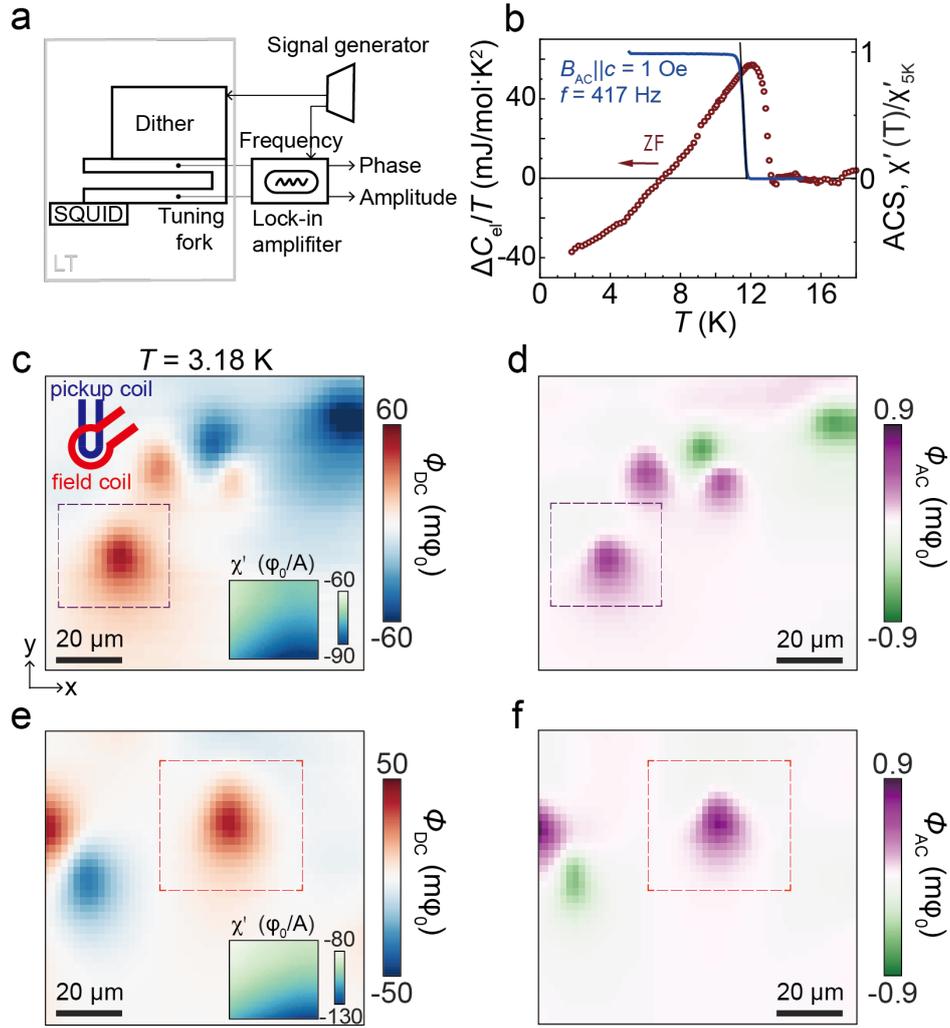

**Figure 1 Imaging vortices in Ba$_{0.23}$K$_{0.77}$Fe$_2$As$_2$ by sSQUID microscopy with a tuning fork. a,** Illustration of the probe setup. The nano-SQUID chip is mounted on a quartz tuning fork which oscillates at its resonant frequency under the drive of a dither piezo. The demodulated alternate-current (AC) flux signal $\Phi_{AC}$ is proportional to the gradient of the magnetic flux along the vibration axis ($z$). It is proportional to the direct-current (DC) magnetic signal $\Phi_{DC}$ when the vibration amplitude is small. **b,** Specific heat (red circles, left axis) and bulk magnetic susceptibility (blue line, right axis) as a function of temperature. **c** and **d,** $\Phi_{DC}$ and $\Phi_{AC}$ images respectively obtained at 3.18 K after a field cooling cycle. Inset: susceptometry image of the area. **e** and **f,** $\Phi_{DC}$ and $\Phi_{AC}$ images respectively of a different area. The integer vortices circled in the boxes appear similar at this temperature.

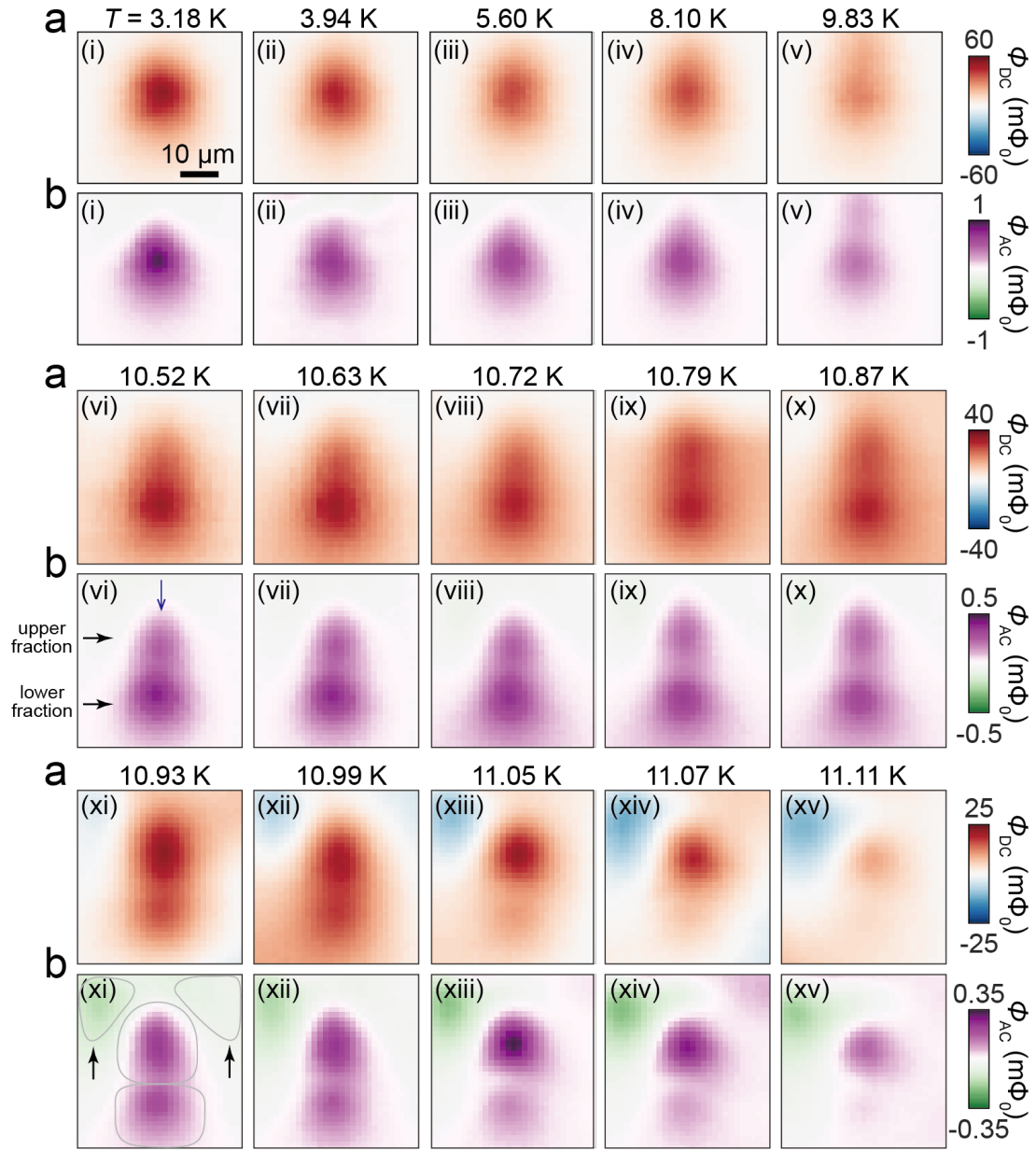

**Figure 2 Splitting of an integer vortex into two fractions with increasing temperature. a** and **b** are the $\Phi_{DC}$ and $\Phi_{AC}$ magnetometry images obtained simultaneously at the corresponding temperatures during a warming sequence. The image obtained at 3.18 K (i) covers a conventional integer vortex (CV) (the red square in Fig. 1f). The upper part of the vortex starts to extrude out from 5.6 K (ii). At 9.83 K (v), a fractional vortex (FV) is clearly visible in $\Phi_{AC}$. The upper FV gains more flux, while the lower one loses intensity with increasing temperature. Note the difference in color scales for the three rows. At elevated temperatures, we see other vortices and antivortices entering, which influences the magnetic flux in the scanning area for $T >$ 10.7 K.

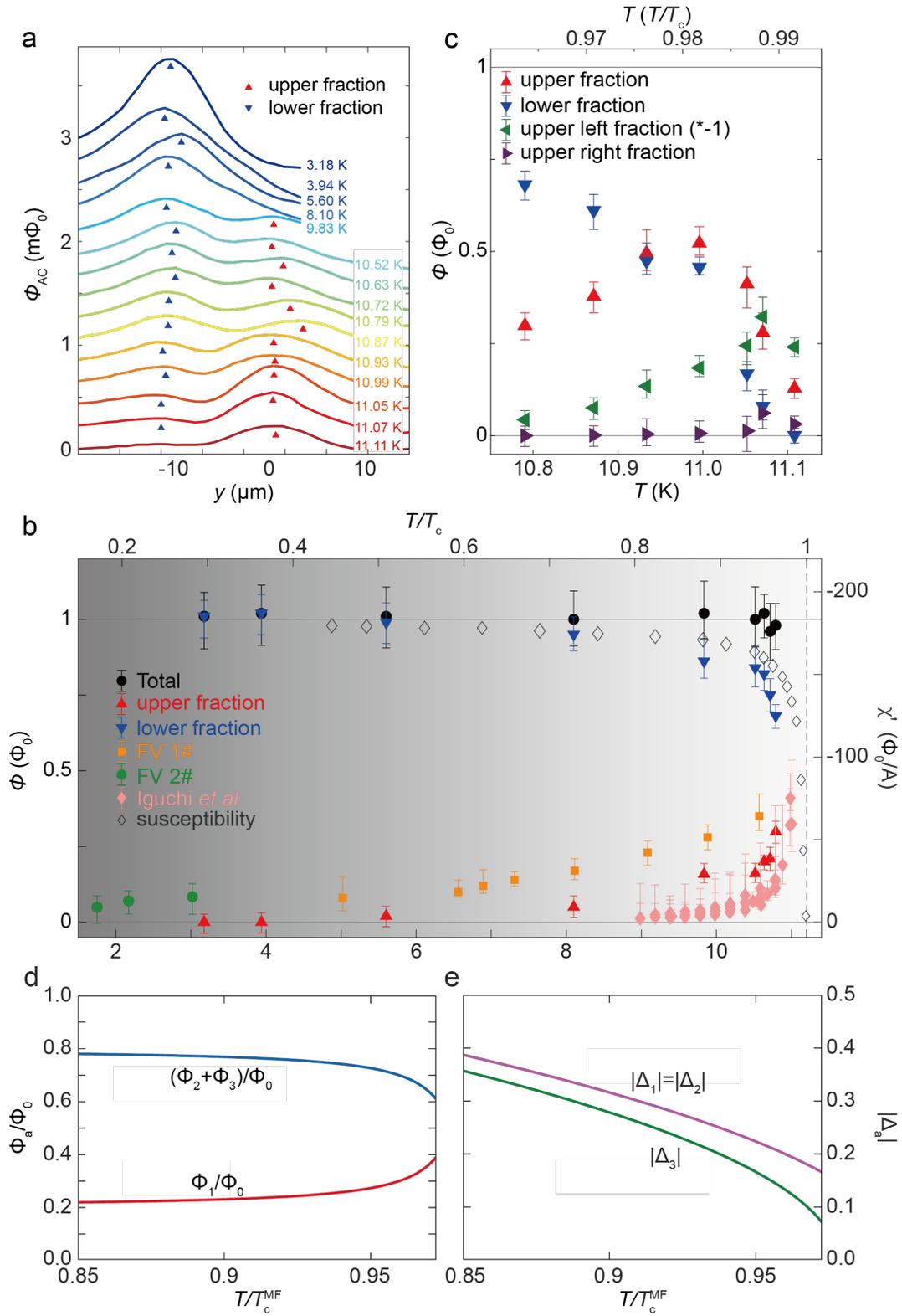

**Figure 3 Flux redistribution between the split fractional vortices. a,** Linecuts of split vortices' $\Phi_{AC}$ in Fig. 2b along their central $y$ axis (Fig. 2b(vi), purple arrow). The centers of the upper and lower fractions are indicted by red and blue triangles, respectively. The changing height of the peaks with increasing temperature shows

transfer of magnetic flux between the two parts of the upper and lower fractions (Fig. 2b(vi), black arrows). **b,** Temperature dependence of the total flux of the FV's (black circles) and that of the upper FV (red square) and lower FV (blue triangle). The data is shown up to 10.79 K to ensure that flux is not strongly affected by other vortices in the scanned area. The total is integrated from the DC flux images, whereas the flux of the FVs is from the ratio of the integrated AC flux of the upper and lower FVs multiplied by their total. The flux of an isolated FV, which is integrated from its DC flux image, is shown as orange squares. It is approximately twice higher than red triangles. The diamagnetic susceptibility of the sample (right axis) is shown in black diamonds and also as the greyscale background. Pink: Flux of FVs from Iguchi *et al*[10]. **c,** flux of the FV's close to $T_C$. The upper left (green) and upper right (purple) refer to the two areas pointed by arrows in Fig. 2b(xi). **d** and **e,** Theoretically calculated flux fractions and gap amplitudes, respectively, based on three-band mean-field microscopic model, as functions of the temperature. The blue curve shows the flux carried by a vortex with phase winding in the first band. The red curve shows the flux carried by a fractional-composite vortex with phase winding in two bands. The details of the derivation, together with the choice of parameters can be found in the supplementary material. The temperature is measured in units of mean-field critical temperature $T_C^{MF}$ (not to be confused with real $T_C$, which is significantly affected by fluctuations in $Ba_{0.23}K_{0.77}Fe_2As_2$).

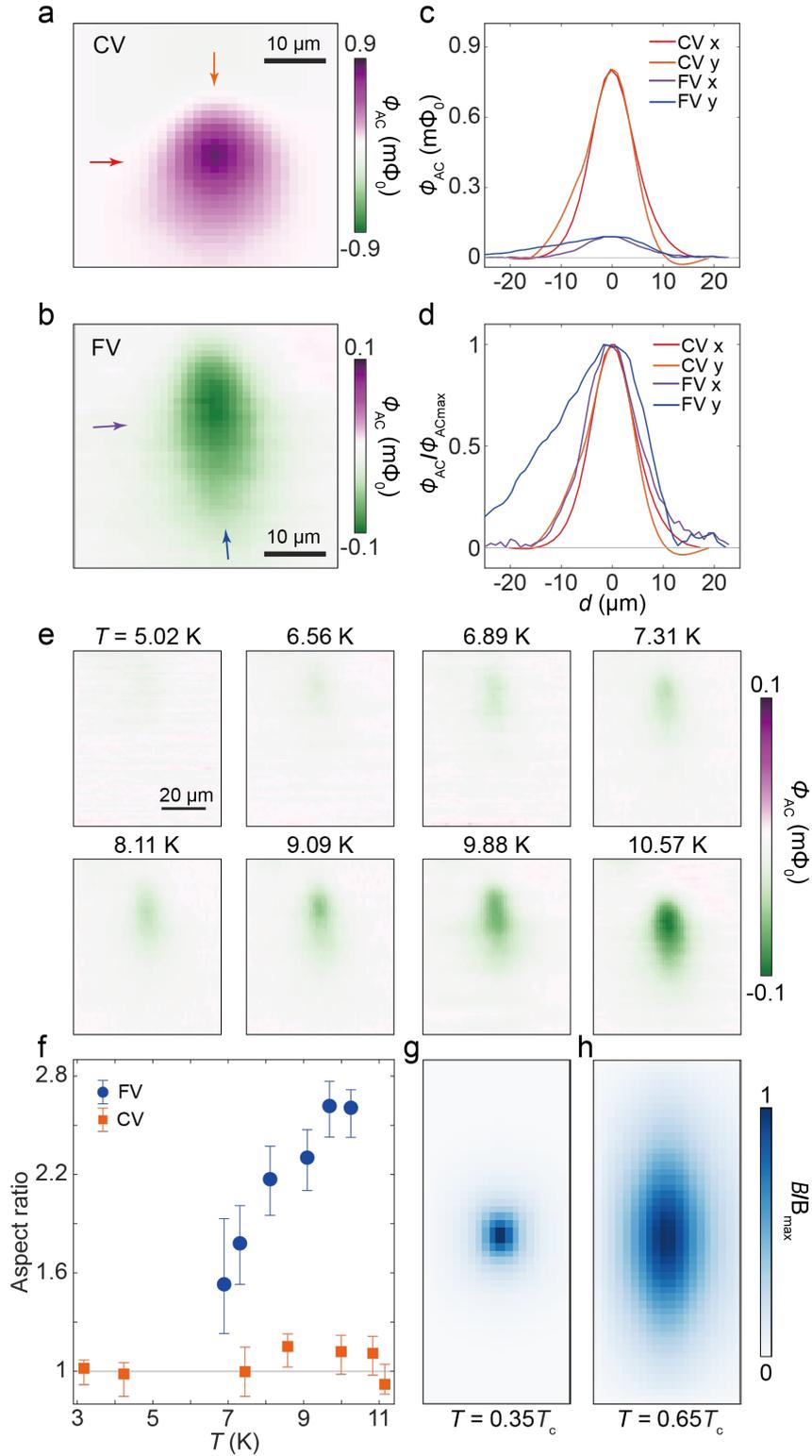

**Figure 4 Elongation of a fractional vortex. a,** $\Phi_{AC}$ image of a CV (purple square area in Fig. 1d), which does not split. **b,** $\Phi_{AC}$ image of a FV at 8.93 K showing an elongated shape. **c,** Linecuts along horizontal (*x*) and vertical (*y*) directions through the centre of the CV (red and orange) and FV (purple and blue) in **a** and **b**, respectively. **d,** Normalized linecuts of **c**. **e,** Temperature dependence $\Phi_{AC}$ of the FV. **f,** Aspect ratios of the FV (blue solid circles) at different temperatures obtained from

**a**. Aspect ratios of a CV in the same temperature range are shown as orange solid squares. **g** and **h,** Normalized magnetic field of simulated elongation of fractional vortex in three-band minimal Bogoliubov-de-Gennes model at $0.35T_C$ and $0.65T_C$, respectively.

# Extended Data Figures

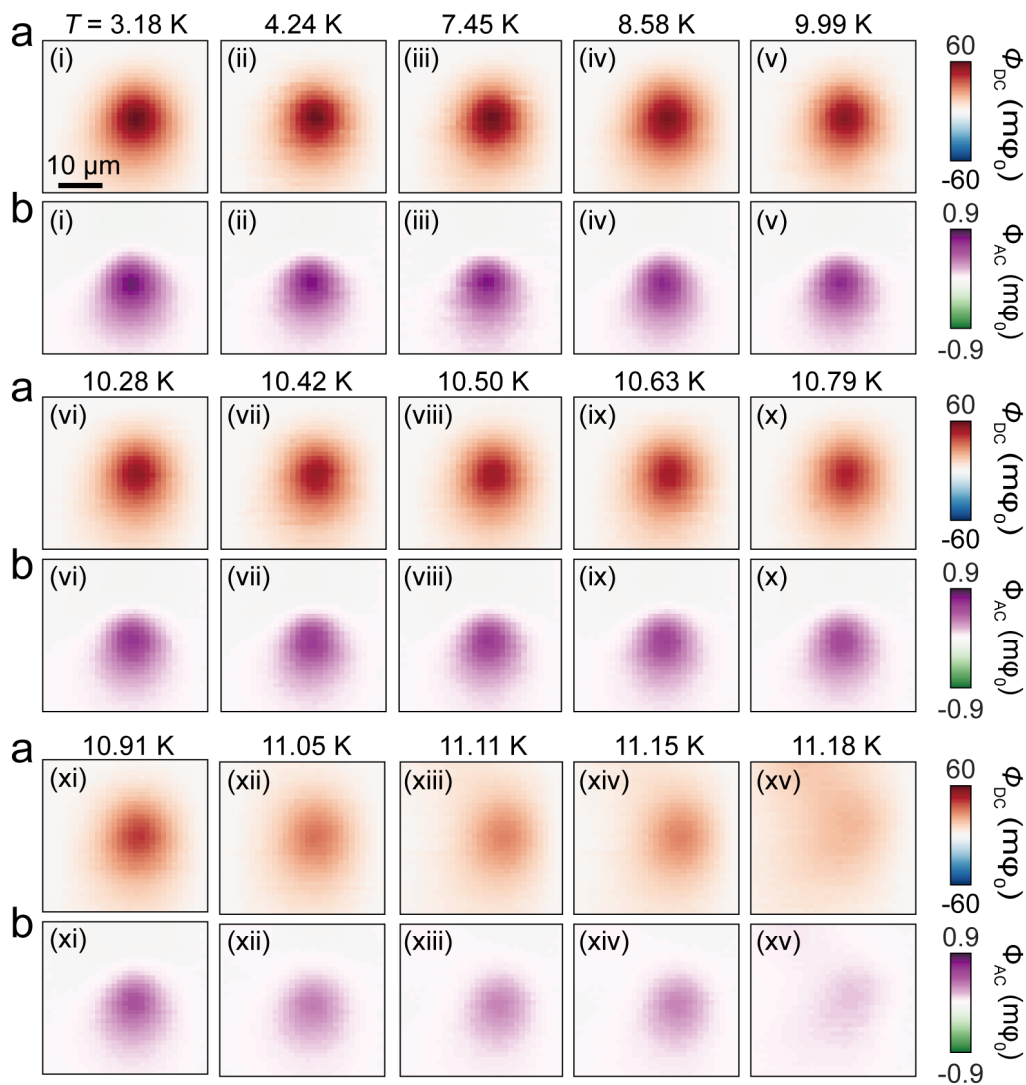

**Extended Data Figure 1 Temperature dependence of a conventional integer vortex.** It is the same vortex as shown in Figs. 1c and d. **a** and **b** are the $\Phi_{DC}$ and $\Phi_{AC}$ magnetometry images obtained simultaneously at the corresponding temperatures following a similar warming sequence as on the FVs shown in Fig. 2.

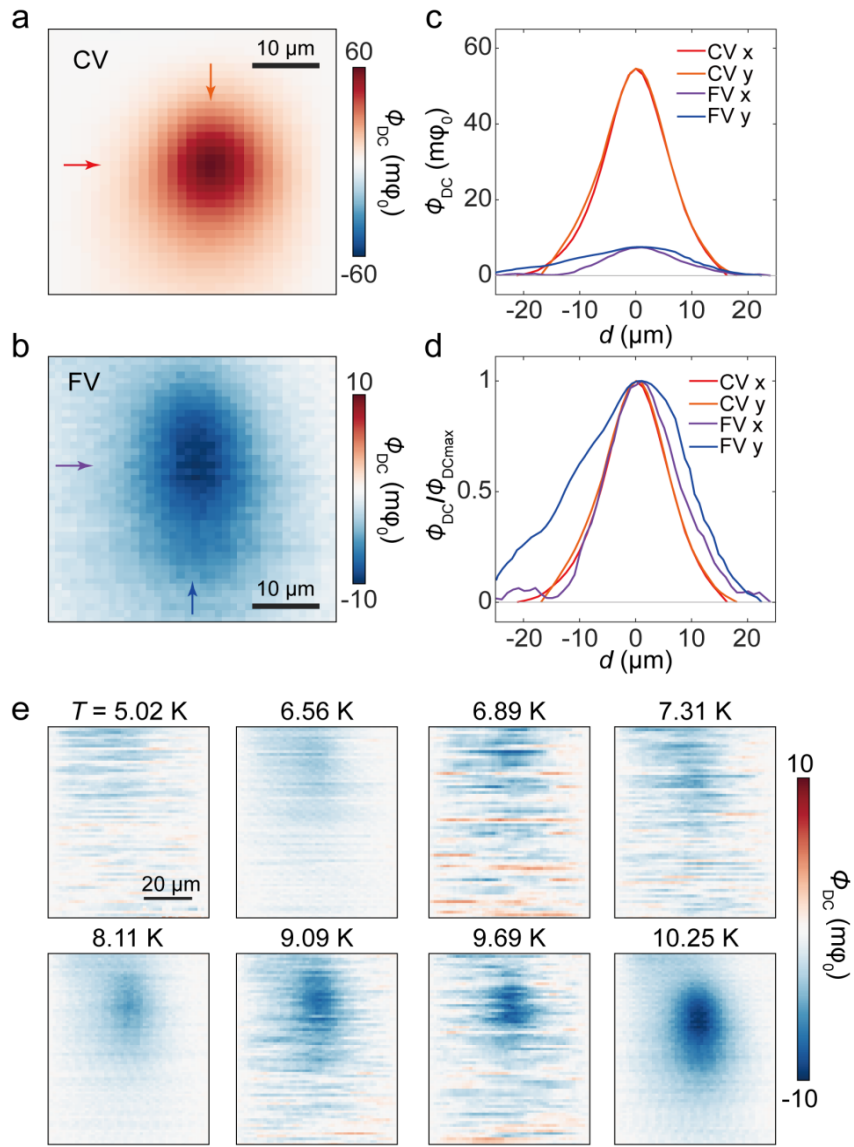

**Extended Data Figure 2 Shape comparison of the conventional and fractional vortex. a** and **b** are the $\Phi_{DC}$ images of the CV and FV, respectively. **c** shows their correspondent linecuts along the horizontal (*x*) and vertical (*y*) directions. **d** shows the normalized linecuts of **c**. **e**, temperature dependence of the $\Phi_{DC}$ images of an isolated CV.

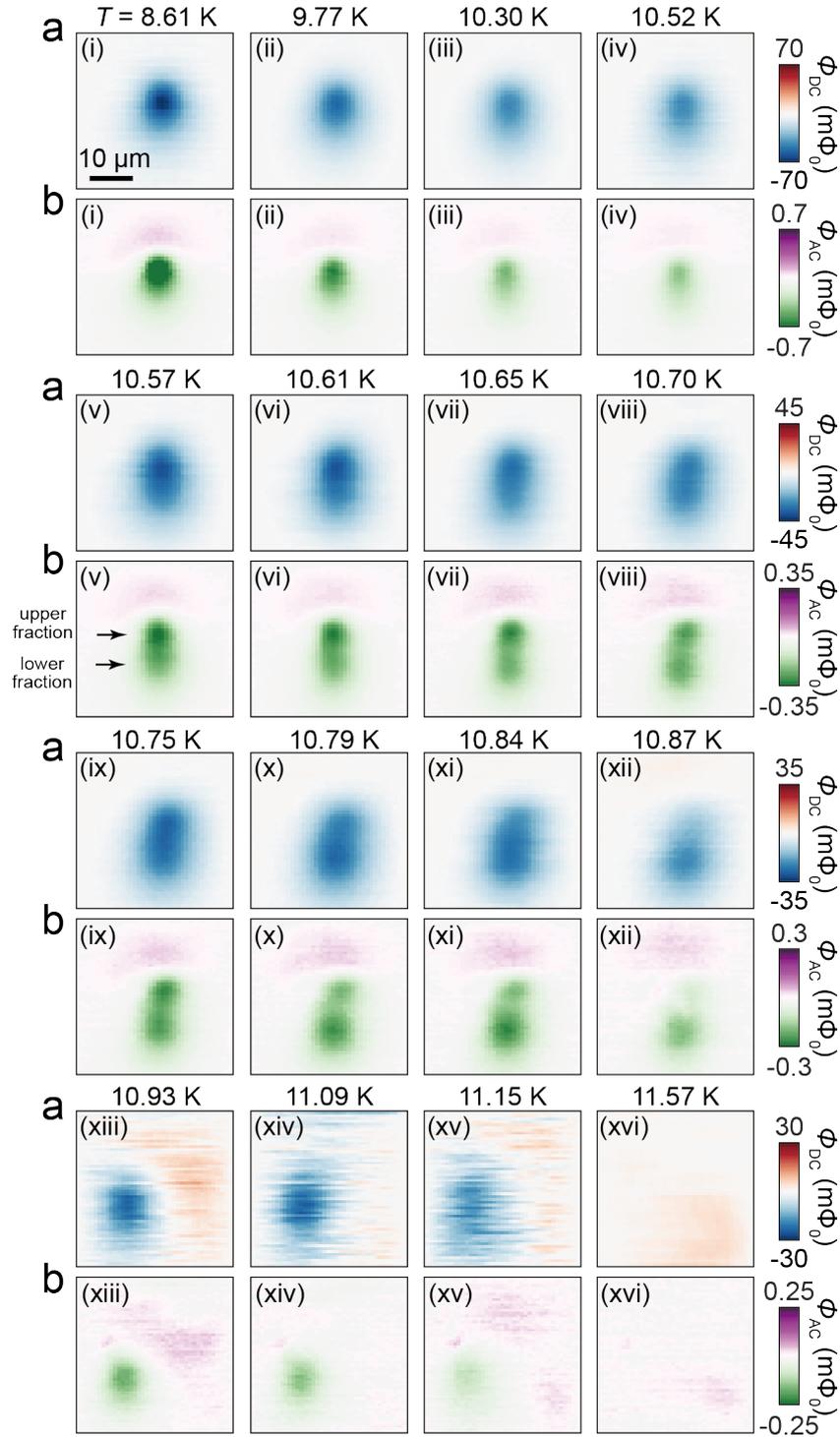

**Extended Data Figure 3 Splitting of another vortex with** $-\Phi_0$. **a** and **b** are the $\Phi_{DC}$ and $\Phi_{AC}$ magnetometry images obtained simultaneously at the corresponding temperatures following a similar warming sequence as on the FVs shown in Fig. 2.

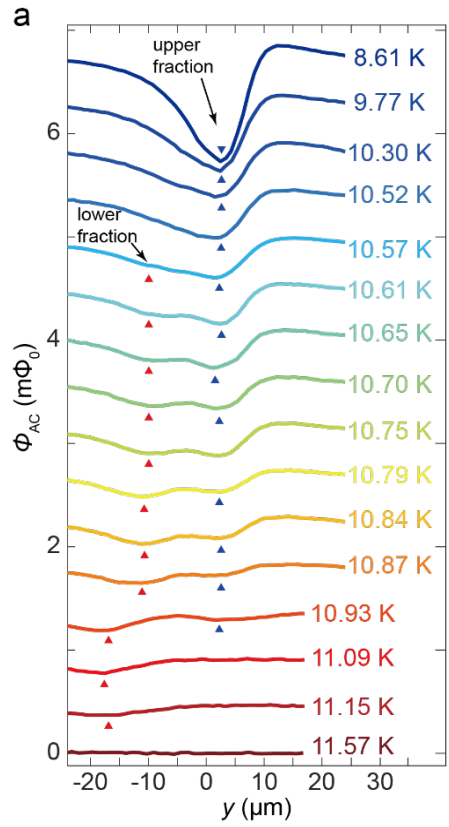

**Extended Data Figure 4 Temperature-dependent peaks of the split fractional vortices with $-\Phi_0$. a,** Linecuts of split vortices' $\Phi_{AC}$ in Extended Data Fig. 3b along their central *y*-axis. The centres of the upper and lower fractions are indicated by red and blue triangles, respectively.

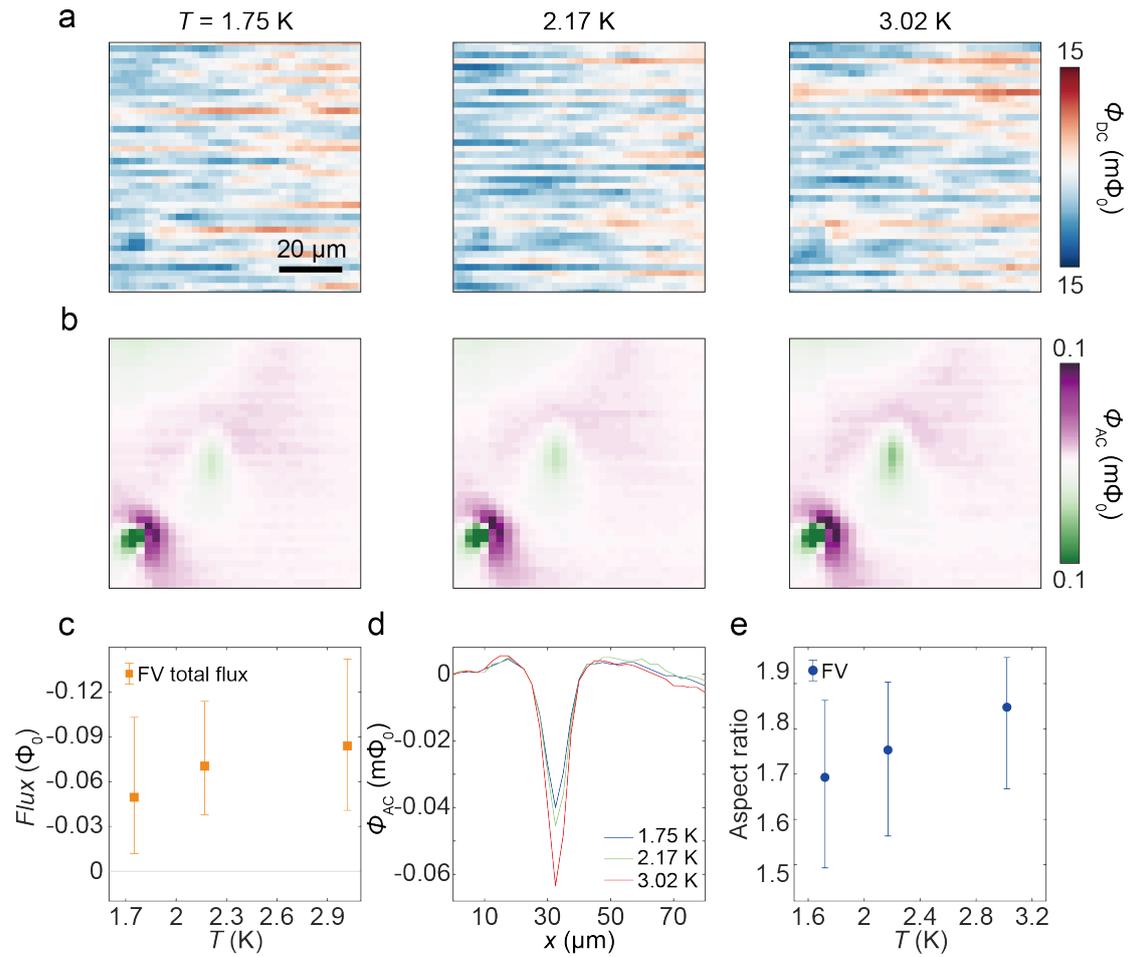

**Extended Data Figure 5 Isolated fractional vortex in the low temperature regime. a** and **b** are the $\Phi_{DC}$ and $\Phi_{AC}$ magnetometry images, respectively, obtained at the temperatures indicated above the panels. The FV is visible in the centre of the $\Phi_{AC}$ images while it is below the noise level in $\Phi_{DC}$. The contrast of the FV is getting bigger with increasing temperature. The stronger feature at the lower left corner of the image is likely from a dipole. **c**, the integrated flux of the vortex at the low-temperature regime obtained from **b**. **d**, Linecuts along the horizontal direction of the FV obtained from **b**. **e**, The aspect ratio of the FV obtained from **b**.